%% file: iclr2025_conference.tex
\documentclass{article} 

\input{math_commands.tex}

\usepackage[
    colorlinks=true,
    linkcolor=linkblue,
    citecolor=linkblue,
    urlcolor=linkblue
]{hyperref}

\usepackage{url}

\usepackage[a4paper,margin=1in]{geometry}
\usepackage[most]{tcolorbox}
\usepackage{xcolor}

\usepackage{newtxtext}

\usepackage{newtxmath}

\usepackage{natbib}
\usepackage{enumitem}
\usepackage{svg}
\usepackage{graphicx}
\usepackage{titlesec}
\usepackage{tabularx}  
\usepackage{booktabs}
\usepackage{multirow}

\usepackage{subcaption}
\usepackage{booktabs}
\usepackage[table]{xcolor}
\usepackage{amsmath}

\newcommand{\alm}[1]{\textcolor{gray}{#1}}

\titlespacing*{\section}
{0pt}{8pt}{4pt}

\titleformat{\section}
{\large\bfseries}
{\thesection}{0.5em}{}

\titleformat{\subsection}
{\normalsize\bfseries}
{\thesubsection}{0.5em}{}

\titlespacing*{\subsection}
{0pt}{4pt}{2pt}

\begin{document}

\definecolor{boxgray}{RGB}{242,243,245}
\definecolor{linkblue}{RGB}{6,69,173}
\definecolor{lightgray}{RGB}{245,245,245}

\begin{center}

\begin{tcolorbox}[
    width=1.0\textwidth,
    colback=lightgray,
    colframe=white,
    arc=5mm,
    boxrule=0.6pt,
    left=6mm,
    right=6mm,
    top=4.5mm,
    bottom=4.5mm
]

{\fontsize{16}{18}\selectfont \bfseries
\includegraphics[
    height=0.45cm,
    trim=250 250 250 250,
    clip
]{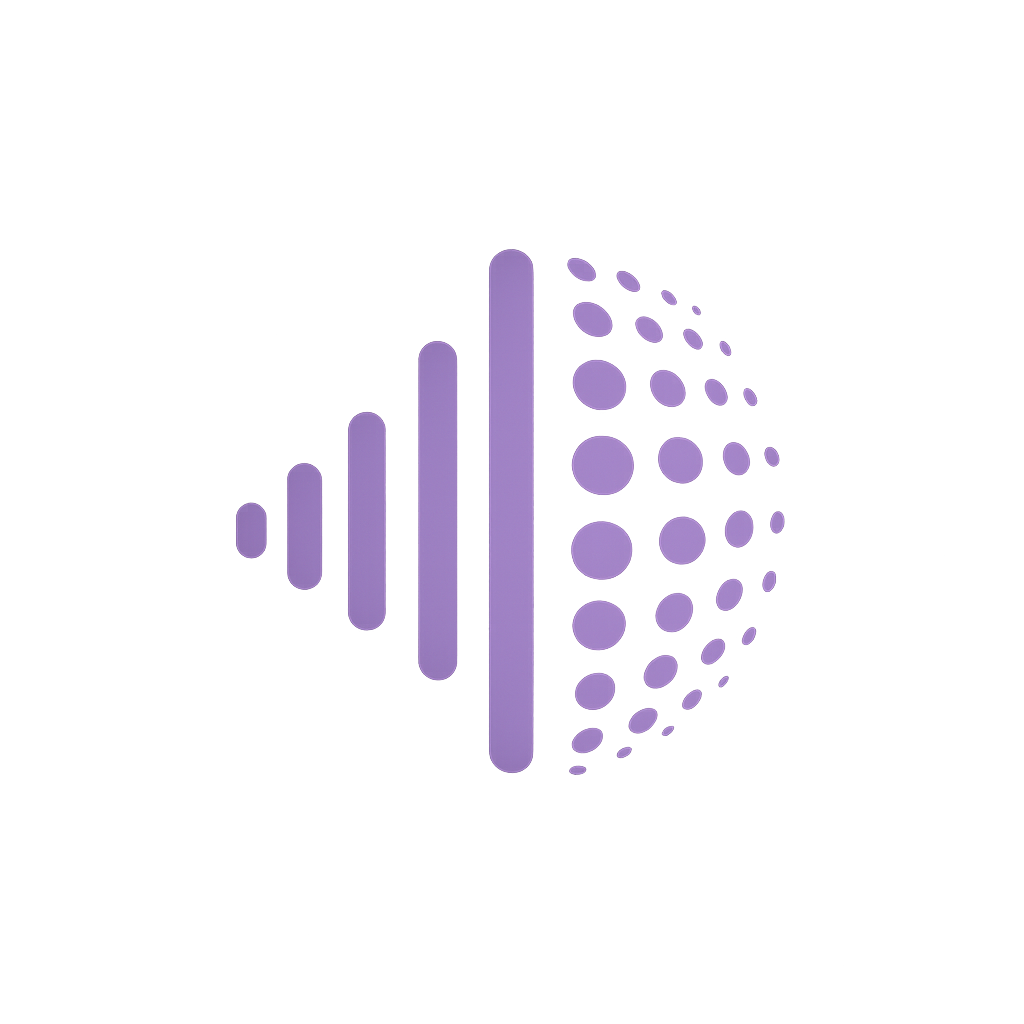}
ALM2Vec: Learning Audio Embeddings 
for Universal \\[0.25em]
Audio Retrieval
with Large Audio-Language Models
}
\vspace{0.3cm}


{\normalsize
\textbf{Fengjie Lu}$^{1}$ \quad
\textbf{Chenang Jiang}$^{1}$ \quad
\textbf{Jiarui Hai}$^{2\dagger}$ \quad
\textbf{Helin Wang}$^{2}$ \quad
\textbf{Aaron Yee}$^{1\ddagger}$

}

\vspace{0.3cm}

{\normalsize
$^{1}$Zhejiang University \quad
$^{2}$Johns Hopkins University \quad
\par}

\vspace{0.15cm}

{\normalsize
$^\dagger$ Research Mentorship \quad
$^\ddagger$ Project Lead
}

\vspace{0.3cm}


{\fontsize{11}{13}\selectfont \bfseries
Abstract:}
Recent advances in language--audio retrieval have been largely driven by contrastive dual-encoder architectures that align audio and text in a shared embedding space. While effective, existing retrieval embeddings are primarily optimized for audio--caption matching, limiting their ability to support diverse retrieval objectives and controllable retrieval behaviors. We present ALM2Vec, a universal audio embedding framework derived from pretrained large audio--language models (LALMs). By transferring the audio understanding, instruction-following, and reasoning capabilities acquired through large-scale multimodal training, ALM2Vec learns a unified embedding space for retrieval across audio domains and task types. Beyond conventional text--audio retrieval, ALM2Vec incorporates natural-language instructions into the embedding process, enabling instruction-aware retrieval for scenarios such as audio question answering and aspect-conditioned retrieval. Experimental results show that ALM2Vec achieves competitive performance on standard audio and speech retrieval benchmarks while exhibiting promising compositional and controllable retrieval capabilities, highlighting its potential as a unified audio embedding model for retrieval across domains, tasks, and user intents.

\vspace{0.3cm}


\noindent
\textbf{Project Page:}
\url{https://caml-labs.github.io/ALM2Vec/}




\end{tcolorbox}

\end{center}



\section{Introduction}

Audio retrieval plays an increasingly important role in applications spanning music, speech, environmental sounds, sound effects, and multimodal media content. As large-scale audio collections continue to grow, retrieval systems require powerful semantic understanding to enable efficient access to relevant content across diverse domains \cite{elizalde2023clap, wu2023large}. Beyond search and recommendation, audio retrieval serves as a key component for data curation, dataset cleaning \cite{hai2024ezaudio, hai2025synsonic, shi2025sam}, and the conditioning and evaluation of generative audio models \cite{liu2023audioldm, ghosal2023text}, driving the demand for robust and generalizable audio embedding models.

Progress in audio retrieval has been largely driven by contrastive dual-encoder frameworks \cite{wu2023large, elizalde2023clap} that learn a shared embedding space for audio and text. While these approaches achieve strong performance on standard text--audio retrieval benchmarks, emerging audio understanding tasks pose new challenges for retrieval systems. Modern applications increasingly involve acoustically complex environments, long-form recordings \cite{drossos2020clotho}, and speech-rich content \cite{hu2026end}, requiring reasoning over temporal structure, multiple sound events, and linguistic information. Meanwhile, retrieval intents are becoming more diverse and instruction-driven. Rather than matching audio based solely on holistic semantic descriptions, users may seek recordings characterized by specific acoustic attributes, sound events, speaker characteristics, or other targeted aspects of audio content. Such objectives extend beyond the global semantic alignment learned by conventional embedding models, demanding finer-grained cross-modal grounding and more controllable retrieval behavior.

Audio large language models (ALLMs) \cite{dinkel2025midashenglm, chu2024qwen2, wu2025step, comanici2025gemini} have demonstrated remarkable audio understanding across a broad range of acoustic domains and task formulations. By combining language reasoning with audio perception, these models can comprehend environmental sounds, music, speech, and long-form recordings while generalizing across tasks such as captioning, question answering, and instruction following. Built upon powerful large language models (LLMs) \cite{yang2025qwen3, team2024gemma, liu2023visual}, ALLMs naturally inherit strong instruction-following and reasoning abilities, enabling them to interpret flexible natural-language instructions and ground them in audio content. These properties make ALLMs particularly well suited for the retrieval scenarios described above. At the same time, several studies \cite{li2026qwen3, zhang2025qwen3, sturua2024jina} have shown that pre-trained LLMs can be effectively adapted into embedding models through representation learning objectives, yielding strong retrieval performance across diverse benchmarks. Building on these developments, we leverage audio large language models to learn a unified retrieval representation for audio. By transferring the audio understanding, instruction-following, and reasoning abilities acquired through large-scale multimodal training, our approach supports retrieval across audio domains, task types, and user intents within a shared embedding space.

In this work, we present
\includegraphics[
    height=0.25cm,
    trim=250 250 250 250,
    clip
]{figs/logo.png}
\textbf{ALM2Vec}. Our main contributions are summarized as follows:

\begin{itemize}[nosep]
\item We release \textbf{ALM2Vec}, an open-source audio embedding model built upon audio large language models, enabling retrieval across diverse audio domains, including sound effects, speech, and music.

\item \textbf{ALM2Vec} achieves state-of-the-art or competitive performance on audio and speech retrieval benchmarks, demonstrating the effectiveness of audio large language models as universal audio embedding learners.

\item Beyond conventional audio retrieval, we investigate instruction-guided embedding extraction for controllable retrieval. Experiments on audio question answering benchmarks and case studies show that \textbf{ALM2Vec} captures user-specified attributes and retrieval intents beyond coarse semantic matching.
\end{itemize}




\section{Method}

\subsection{Model Architecture}

\begin{figure*}
  \centering
  \includegraphics[width=0.90\linewidth]{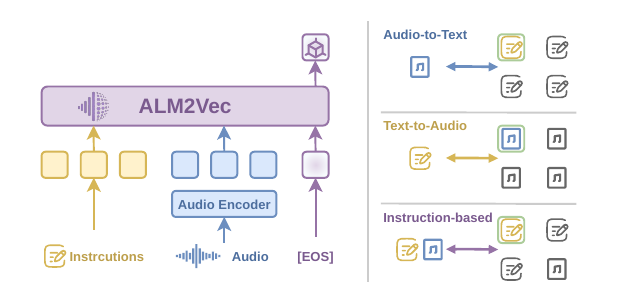}
\vspace{-0.5cm}
\caption{Overview of ALM2Vec framework and retrieval tasks.}
\label{fig:overview}
\end{figure*}

The ALM2Vec model is built upon the pre-trained MiDashengLM model \cite{dinkel2025midashenglm}. MiDashengLM comprises a mel-spectrogram-based audio transformer encoder \cite{dinkel2024scaling} and a Qwen2.5-based language model \cite{Qwen2.5-Omni}, enabling unified processing of text and diverse audio modalities, including speech, music, and general acoustic events. Pre-trained on large-scale audio-text data with audio understanding and instruction-following objectives, the model provides rich acoustic and semantic representations. These pre-trained capabilities serve as a strong foundation for transfer learning, allowing ALM2Vec to efficiently adapt MiDashengLM for universal audio representation learning. Furthermore, its ability to process long-form audio makes it suitable for a wide range of audio understanding tasks.


As illustrated in Figure~\ref{fig:overview}, ALM2Vec encodes each input using the ALLM-based backbone. An input may consist of text, audio, or a combination of both, accompanied by an instruction. The hidden state corresponding to the final \texttt{[EOS]} token is used as a global representation that captures the semantics of the instruction and the input content. This representation is then projected into a fixed-dimensional embedding space to obtain the final embedding.

By mapping text, audio, and multimodal inputs into a unified embedding space, ALM2Vec enables direct comparison across different modalities. This unified representation facilitates cross-modal retrieval and allows the model to learn semantic relationships between heterogeneous inputs. Furthermore, the instruction-aware nature of the backbone enables the resulting embeddings to capture task-relevant aspects of audio, including semantic content, acoustic characteristics, speaker attributes, and audio quality.

\subsection{Learning Objective}

ALM2Vec is trained using a contrastive learning objective for retrieval \cite{radford2021learning}. Given a mini-batch of $N$ query-document pairs, ${(q_i, d_i)}_{i=1}^{N}$, where each query $q_i$ is associated with its corresponding relevant document $d_i$, both inputs are encoded using the shared embedding model. Following the embedding extraction procedure described above, the resulting representations are projected and L2-normalized to obtain query and document embeddings:

\begin{equation}
\mathbf{z}^{q}_{i}
=
\frac{f_{\theta}(q_i)}
{\|f_{\theta}(q_i)\|_2},
\qquad
\mathbf{z}^{d}_{i}
=
\frac{f_{\theta}(d_i)}
{\|f_{\theta}(d_i)\|_2},
\end{equation}

where $f_{\theta}(\cdot)$ denotes the shared model and $\|\cdot\|_2$ denotes L2 normalization. The similarity between a query embedding and a document embedding is computed using scaled cosine similarity:

\begin{equation}
s_{ij}
=
\frac{
(\mathbf{z}^{q}_{i})^\top \mathbf{z}^{d}_{j}
}{\tau},
\end{equation}

where $\tau$ is a learnable temperature parameter that controls the sharpness of the similarity distribution. For each query $q_i$, the paired document $d_i$ is treated as a positive example, while all other documents in the mini-batch act as negative examples. Likewise, for each document $d_i$, all non-matching queries are treated as negatives. This in-batch negative sampling strategy enables efficient contrastive training without requiring additional negative examples.

Following standard bidirectional contrastive learning, the objective optimizes retrieval performance in both query-to-document and document-to-query directions:

\begin{equation}
\mathcal{L}
=
\mathcal{L}_{q\rightarrow d}
+
\mathcal{L}_{d\rightarrow q}
=
-\frac{1}{N}
\sum_{i=1}^{N}
\log
\frac{\exp(s_{ii})}
{\sum_{j=1}^{N}\exp(s_{ij})}
-\frac{1}{N}
\sum_{i=1}^{N}
\log
\frac{\exp(s_{ii})}
{\sum_{j=1}^{N}\exp(s_{ji})}.
\end{equation}

The query-to-document loss $\mathcal{L}{q\rightarrow d}$ treats each query as an anchor and encourages its corresponding document to achieve the highest similarity among all documents in the mini-batch. Symmetrically, the document-to-query loss $\mathcal{L}{d\rightarrow q}$ treats each document as an anchor and encourages its paired query to achieve the highest similarity among all queries in the mini-batch. Optimizing both directions improves retrieval performance and learns a shared embedding space where semantically related query-document pairs are positioned close together while unrelated pairs are pushed apart.

\subsection{Training Details}

Leveraging its ability to process audio and text within a unified sequence, ALM2Vec is trained on both audio captioning datasets, such as AudioCaps \cite{kim2019audiocaps} and Clotho \cite{drossos2020clotho}, and audio question answering (QA) datasets \cite{goel2025audioflamingo3advancing, ghosh2025audio}. During training, audio QA samples are reformulated as retrieval pairs, where the query consists of the question and associated audio, and the answer serves as the document. This training setup exposes the model to a diverse range of audio, speech, and music understanding tasks, ranging from general audio summarization to detailed understanding of acoustic events, spoken content, speaker characteristics, and musical attributes.

Training consists of two stages: pretraining and fine-tuning. During pretraining, audio inputs are limited to 15 seconds, and the model is trained for 4,000 steps with a global batch size of 256 across 8 NVIDIA PRO 6000 GPUs. During fine-tuning, the maximum audio length is increased to 30 seconds, and training continues for 2,000 steps with an effective batch size of 64 on 2 GPUs. To better align the model with audio retrieval benchmarks, the fine-tuning stage increases the proportion of summarization-style data while retaining general audio QA samples to mitigate catastrophic forgetting. Furthermore, to improve the stability of contrastive learning and reduce the impact of false negatives, in-batch negative pairs with high similarity scores are masked during loss computation.

Throughout training, the Dasheng audio encoder remains frozen, while the remaining components are adapted using LoRA applied to the query, key, and value projection layers of the language model. The LoRA configuration uses rank $r=16$, scaling factor $\alpha=32$, and a dropout rate of 0.05. Optimization is performed using AdamW with a weight decay of $10^{-3}$ and a warmup-cosine learning-rate schedule, where the learning rate linearly warms up over the first 500 steps to a peak value of $10^{-4}$ before decaying to $10^{-5}$.

\section{Evaluation}

\subsection{Evaluation Tasks}

\textbf{Audio-Text Retrieval.}
Audio-text retrieval performance is evaluated on the widely used audio captioning datasets \cite{kim2019audiocaps, drossos2020clotho} under both text-to-audio and audio-to-text settings. This task measures the alignment between audio signals and their corresponding textual descriptions by retrieving the correct audio given a text query, or vice versa, where Recall@K is used as the evaluation metric. Open-source CLAP-based models \cite{elizalde2023clap, wu2023large, mei2024wavcaps}, along with the multimodal language model–based embedding model jina-embeddings-v5-omni \cite{honicke2026jina}, are included for comparison.

\textbf{Speech-Text Retrieval.}
In addition to audio--text retrieval, which primarily focuses on acoustic event descriptions, we further evaluate semantic speech--text retrieval on LibriSQA \cite{zhao2024librisqa}. This task requires retrieving the correct textual question or answer given a speech query, or the corresponding speech utterance given a text query, thereby assessing whether the model captures linguistic and semantic information in speech. We report Recall@K for both speech-to-text and text-to-speech retrieval. As baselines, we include CLSR \cite{hu2026end}, an end-to-end speech-language retrieval model, and a cascaded approach that first transcribes speech using Whisper ASR \cite{radford2023robust} and then encodes the resulting text with a BGE-based retriever \cite{llm_embedder}. We also include CLAP \cite{wu2023large} for reference, noting that it is not directly comparable because it is trained on audio-caption data rather than semantic speech understanding tasks; its results nevertheless illustrate a capability that is absent from conventional audio embedding models.

\textbf{Audio Question Answering.}
To further evaluate the model's audio understanding capabilities, we assess performance on audio question--answer selection tasks. In this setting, the query consists of a text question paired with an audio input, while the candidate pool contains multiple textual answer choices. The query embedding is obtained by encoding the audio--question pair, and each answer choice is encoded independently as a candidate embedding. The model must identify the correct answer by retrieving the most relevant candidate for the audio--text query. Experiments are conducted on the MMAU-Mini benchmark \cite{sakshi2025mmau}, and performance is measured using accuracy. We include several state-of-the-art audio-language models \cite{goel2025audioflamingo3advancing, Qwen2.5-Omni, comanici2025gemini, hurst2024gpt} as reference points for comparison.

\subsection{Evaluation Results}

\begin{table*}[t]
\caption{Text-audio retrieval results on the AudioCaps and Clotho test sets. }
\label{tab:retrieval_results}
\footnotesize
\centering
\begin{tabular}{l|ccc|ccc||ccc|ccc}
\toprule
& \multicolumn{6}{c||}{AudioCaps}
& \multicolumn{6}{c}{Clotho} \\
\cline{2-13}
Method
& \multicolumn{3}{c|}{Text-to-Audio}
& \multicolumn{3}{c||}{Audio-to-Text}
& \multicolumn{3}{c|}{Text-to-Audio}
& \multicolumn{3}{c}{Audio-to-Text} \\
\cline{2-13}
& R@1 & R@5 & R@10
& R@1 & R@5 & R@10
& R@1 & R@5 & R@10
& R@1 & R@5 & R@10 \\
\midrule

LAION-CLAP
& 36.1 & 71.8 & 83.9
& 46.8 & \underline{82.9} & \underline{90.7}
& 16.1 & 38.3 & 51.1
& 22.7 & 48.5 & 60.8
\\

MS-CLAP
& 15.4 & 47.2 & 64.5   
& 32.0 & 66.0 & 79.2   
& 15.6 & 38.9 & 51.4
& 22.1 & 48.9 & 62.0
\\

WavCaps-CLAP-PT
& 39.7 & 74.5 & 86.1
& 51.7 & 82.3 & 90.6
& 19.5 & 45.2 & 58.2
& 23.4 & 50.9 & 63.4
\\

WavCaps-CLAP-FT
& \underline{42.2} & \underline{76.5} & \underline{87.1}
& \underline{54.6} & \textbf{85.2} & \textbf{92.4}
& \underline{19.7} & \underline{45.7} & \underline{59.4}
& \underline{26.9} & \underline{52.6} & \underline{64.9}
\\

JINA-Embed.-v5
& 20.4 & 50.3 & 64.4   
& 23.1 & 52.7 & 67.2   
& 9.2  & 23.9 & 35.0   
& 10.5 & 24.7 & 34.3   
\\

\midrule

\rowcolor{cyan!4}
\textbf{ALM2Vec-PT}
& 40.0 & 74.5 & 85.9
& 43.8 & 74.3 & 86.5
& 19.2 & 43.4 & 55.7
& 17.9 & 39.4 & 52.2
\\
\rowcolor{cyan!4}
\textbf{ALM2Vec-FT}
& \textbf{43.2} & \textbf{78.0} & \textbf{87.8}
& \textbf{55.5} & 80.0 & 88.2
& \textbf{24.8} & \textbf{52.9} & \textbf{65.8}
& \textbf{27.9} & \textbf{52.7} & \textbf{66.3}
\\

\bottomrule
\end{tabular}
\end{table*}

\begin{table*}[t]
\centering
\footnotesize

\begin{minipage}[t]{0.54\textwidth}
\centering
\captionof{table}{LibriSQA retrieval results.}
\label{tab:librisqa}

\begin{tabular}{l|ccc|ccc}
\toprule
\multirow{2}{*}{Method}
& \multicolumn{3}{c|}{Text-to-Speech}
& \multicolumn{3}{c}{Speech-to-Text} \\
\cline{2-7}
& R@1 & R@5 & R@10
& R@1 & R@5 & R@10 \\
\midrule


\alm{LAION-CLAP}
& \alm{0.0} & \alm{0.1} & \alm{0.8}
& \alm{0.1} & \alm{0.2} & \alm{0.6} \\

Whisper+BGE
& 83.7 & 93.3 & 94.9
& 85.2 & 93.4 & 95.3 \\

CLSR
& \textbf{85.0} & \underline{93.4} & \underline{95.0}
& \underline{85.5} & \underline{94.0} & \underline{95.6} \\
\midrule

\rowcolor{cyan!4}
\textbf{ALM2Vec-PT}
& 43.7 & 64.5 & 72.8
& 11.2 & 24.9 & 34.1 \\

\rowcolor{cyan!4}
\textbf{ALM2Vec-FT}
& \underline{84.7} & \textbf{94.1} & \textbf{95.8}
& \textbf{86.0} & \textbf{95.2} & \textbf{97.2} \\

\bottomrule
\end{tabular}
\end{minipage}
\hfill
\begin{minipage}[t]{0.44\textwidth}
\centering

\captionof{table}{Performance on the MMAU-mini.}
\label{tab:mmau}

\begin{tabular}{l|c|ccc}
\toprule
Method & Overall & Music & Sound & Speech \\
\midrule

\alm{GPT-4o Audio} &
\alm{60.8} &
\alm{63.2} &
\alm{64.6} &
\alm{56.3} \\

\alm{Gemini 2.5 Pro} &
\alm{\underline{71.6}} &
\alm{\underline{75.1}} &
\alm{71.5} &
\alm{68.3} \\

\alm{Qwen2.5-Omni} &
\alm{71.5} &
\alm{65.9} &
\alm{\underline{78.1}} &
\alm{\underline{70.6}} \\

\alm{Audio Flamingo 3} &
\alm{\textbf{73.1}} &
\alm{\textbf{76.9}} &
\alm{66.1} &
\alm{\textbf{73.9}} \\
\midrule

\rowcolor{cyan!6}
\textbf{ALM2Vec-PT} &
66.3 &
62.3 &
\textbf{78.7} &
58.0 \\

\rowcolor{cyan!6}
\textbf{ALM2Vec-FT} &
63.0 &
61.7 &
74.8 &
52.6 \\

\bottomrule
\end{tabular}
\end{minipage}

\label{tab:downstream}
\end{table*}

\textbf{Audio-Text Retrieval.}
Table~\ref{tab:retrieval_results} shows that ALM2Vec consistently benefits from retrieval fine-tuning and achieves competitive performance relative to strong CLAP-based baselines on both AudioCaps and Clotho. While the improvements on AudioCaps are modest, ALM2Vec yields substantially larger gains on Clotho. Because Clotho contains longer recordings with richer acoustic context and multiple overlapping sound events, successful retrieval requires modeling long-range temporal dependencies and integrating information across extended audio segments. The stronger performance on Clotho suggests that the proposed architecture is particularly effective at capturing long-duration audio semantics compared with conventional CLAP models.

\textbf{Speech-Text Retrieval.}
As shown in Table~\ref{tab:librisqa}, the pretrained model achieves only modest performance on semantic speech--text retrieval, indicating that it acquires a certain degree of speech content understanding despite not being explicitly optimized for this task. Retrieval fine-tuning further improves performance substantially, resulting in the best overall results among all embedding-based baselines. Notably, the fine-tuned model also surpasses the cascaded Whisper ASR + BGE retrieval pipeline, demonstrating that the learned representations can capture semantic information directly from speech without relying on an intermediate transcription stage. These results highlight the strong transferability of the pretrained audio--language embedding space for semantic speech understanding and downstream speech tasks.

\textbf{Audio Question Answering.}
Table~\ref{tab:mmau} evaluates whether the learned embeddings support instruction-conditioned audio understanding. Unlike standard retrieval tasks that rely on generic audio representations, MMAU-mini requires the model to extract query-specific information from audio conditioned on a textual question. Despite being trained primarily for retrieval, ALM2Vec achieves competitive performance relative to several large audio-language models, suggesting that the learned embedding space captures question-dependent audio information rather than merely generic audio summaries. Interestingly, retrieval fine-tuning results in a slight performance drop, indicating a potential trade-off between optimizing representations for retrieval and preserving the broader reasoning capabilities required for instruction-conditioned audio understanding.

\subsection{Instruction Following Analysis}

\begin{figure*}[t]
    \centering
    \includegraphics[width=0.95\linewidth]{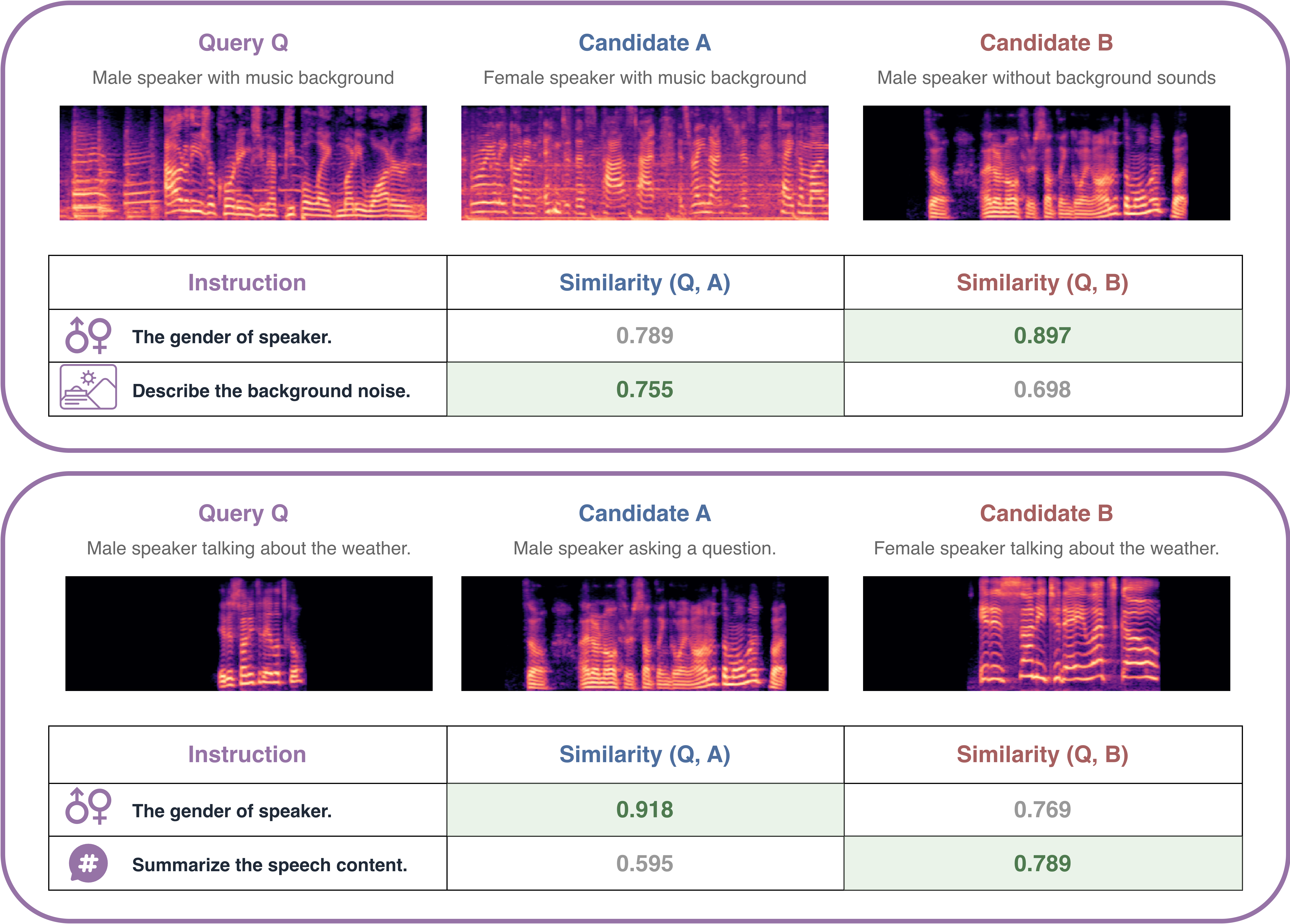}
    \caption{Case studies of instruction-guided audio retrieval. Changing the retrieval instruction alters the retrieved target despite using the same query and candidate audios, demonstrating controllable audio retrieval.}
    \label{fig:case_study}
\end{figure*}

To better understand how the learned embedding space responds to instructions, we present retrieval examples constructed from \emph{confused triplets}, each consisting of a query audio, a semantically correct target, and a hard negative. The hard negative is intentionally selected to share dominant acoustic characteristics with the query while differing in the attribute specified by the instruction. Such examples are particularly challenging because models that rely primarily on generic audio summarization or global audio similarity tend to assign high similarity scores to these hard negatives. As illustrated in Figure~\ref{fig:case_study}, the retrieved result varies according to the instruction, even when competing candidates share strong acoustic similarities with the query. Across examples involving speaker identity, speech content, background sounds, and environmental context, the model consistently emphasizes the instruction-relevant attribute rather than relying on overall audio similarity. These qualitative results suggest that the learned embedding space supports query-dependent audio representations that can selectively attend to different aspects of an audio signal based on the accompanying instruction. Additional examples, together with corresponding audio samples, are available on the \href{https://caml-labs.github.io/ALM2Vec/}{ALM2Vec demo page}.

\section{Conclusion}

We introduced ALM2Vec, a universal audio embedding framework derived from pretrained large audio--language models. ALM2Vec learns a unified embedding space for retrieval across diverse audio domains and tasks while enabling instruction-aware retrieval through natural-language conditioning. Experimental results demonstrate competitive performance on audio--text and speech--text retrieval benchmarks, as well as the ability to produce instruction-guided audio representations. Future work will explore fine-grained cross-modal reranking and broader downstream applications of ALM2Vec, such as audio generation evaluation.

\bibliography{iclr2025_conference}
\bibliographystyle{unsrt}

\end{document}

%% file: math_commands.tex
\usepackage{amsmath,amsfonts,bm}

\def\eqref#1{equation~\ref{#1}}

\def\1{\bm{1}}

\DeclareMathAlphabet{\mathsfit}{\encodingdefault}{\sfdefault}{m}{sl}
\SetMathAlphabet{\mathsfit}{bold}{\encodingdefault}{\sfdefault}{bx}{n}

